\documentclass[%
 reprint,
 amsmath,amssymb,
 aps,
]{revtex4-1}

\usepackage{graphicx}
\usepackage{dcolumn}
\usepackage{bm}
\usepackage{url}
\usepackage{amssymb}
\usepackage{wasysym}
\usepackage[stable]{footmisc}
\usepackage{slashbox}
\usepackage{longtable}
\usepackage{amsmath}
\usepackage{tabularx}
\usepackage{float}

\begin{document}
\preprint{APS/123-QED}
\title{Near-Field Microwave Magnetic Nanoscopy of Superconducting Radio Frequency Cavity Materials}

\author{Tamin Tai$^{1,2}$}
\author{Behnood G. Ghamsari$^{2}$}
\author{Thomas R. Bieler$^{3}$}
\author{Teng Tan$^{4}$}
\author{X. X. Xi$^{4}$}
\author{Steven M. Anlage$^{1,2}$}

\affiliation{$^{1}$Department of Electrical and Computer
Engineering, University of Maryland, College Park, MD 20742-3285,
USA}

\affiliation{$^{2}$Department of Physics, Center for Nanophysics
and Advanced Materials, University of Maryland, College Park, MD  20742-4111, USA}

\affiliation{$^{3}$Chemical Engineering and Materials Science, Michigan State University, East Lansing, MI 48824, USA}

\affiliation{$^{4}$Physics Department, Temple University,
Philadelphia 19122, USA}

\date{\today}
\newcommand{\squeezeup}{\vspace{-2.5mm}}

\begin{abstract}
A localized measurement of the RF critical field on superconducting radio frequency (SRF) cavity materials is a key step to identify specific defects that produce quenches of SRF cavities. Two new measurements are performed to demonstrate these capabilities with a novel near-field scanning probe microwave microscope. The first is a third harmonic nonlinear measurement on a high Residual-Resistance-Ratio bulk Nb sample showing strong localized nonlinear response for the first time, with surface RF magnetic field $B_{surface} \sim 10^2$ mT. The second is a raster scanned harmonic response image on a high quality MgB$_2$ thin film demonstrating a quench defect-free surface over large areas.
\end{abstract}

\maketitle

Materials issues limit the ability to mass-produce bulk Nb Superconducting Radio Frequency (SRF)
cavities with consistent high accelerating gradient performance. Much debate regarding this inconsistent performance
has concluded that certain types of defects on the Nb cavity surface
behave as a source of quenching in high RF magnetic fields \cite{Ciovati2006}\cite{Ciovati2010}. Quench occurs when the
superconductor returns to the normal state in the presence of strong fields, thus limiting the utility of the entire SRF cavity. In addition it is widely observed that the quality factor (Q) of SRF cavities falls dramatically with increasing surface RF magnetic field, particularly as the quench field is approached \cite{Padamsee}. A detailed microscopic understanding of this Q drop is lacking but is most likely related to the inevitable defects on the inner surface of bulk Nb cavities \cite{Ciovati2006}\cite{Ciovati2010}. These defects may be either non-superconducting or have lower critical temperature ($T_c$), making them sources of dissipation which interrupt the superconducting current flow \cite{Halbritter1995}. Unfortunately, it is difficult to totally remove all of these defects even after sophisticated physical and chemical treatments. Because different defects each have their own quench limit, not all of the defects behave as sources of
quenching under given cavity operation conditions. Therefore it is necessary to quantitatively understand the relation between physical defects and their individual electromagnetic properties such as local critical field.

Currently, in order to isolate specific defects in SRF cavities, most measurements utilize a complete accelerator cavity \cite{Liepe} and image the quenching hot spots at high accelerating gradient via a thermometer array attached to the outside wall of the cavity \cite{Ciovati2010}. Such measurements reveal the locations of quench sites only to within a few millimeters, far larger than the defect itself, making their unique identification difficult. Microscopic techniques have not been used in the past because traditionally it is difficult to create strong and localized RF magnetic fields at the cavity operating frequency (several GHz) \cite{Wellstood}\cite{Kikuchi}. Our idea is to look at the active surface of SRF cavities with a very intense and localized magnetic field (on the order of the thermodynamic critical field of bulk Nb), and directly measure the local electromagnetic response in the cavity operating regime at low temperature \cite{Anlage}. Our method is to integrate a magnetic write head from a longitudinal recording hard drive \cite{ReadRiteCorp} into a near-field microwave microscope \cite{Tai2011}\cite{Tai2013} and directly measure the local nonlinear response in the multi-GHz frequency regime at cryogenic temperatures. The microscope can study bulk samples from Nb cavities or any material of interest to the SRF community, such as MgB$_2$ thin films \cite{Tajima}\cite{Xi2009}. Defective regions which create quenches in SRF cavities are expected to be strong sources of nonlinear response. Up to this point localized nonlinear microwave response has only been measured from superconducting thin films \cite{Tai2011}\cite{Tai2013}\cite{S. C. Lee1} \cite{S. C. Lee2}\cite{S. C. Lee3}\cite{S. C. Lee4}\cite{D. Miercia}\cite{Tai2012}. Here we show for the first time a strong localized nonlinear response from the technologically relevant bulk Nb material.
\begin{figure}
   \centering
   \includegraphics*[width=3.4 in]{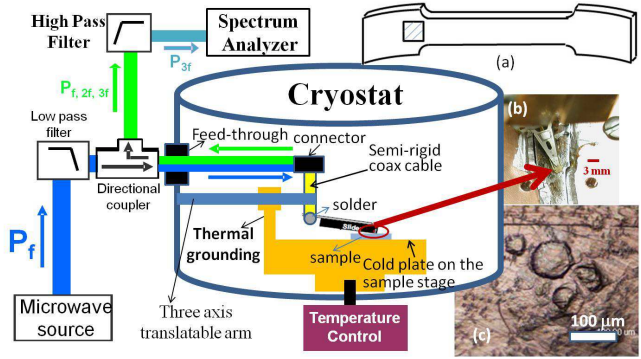}
   \begin{quote}
   \caption{Set up of the $P_{3f}$ measurement in nonlinear microwave microscopy \cite{Tai2013}\cite{S. C. Lee1}\cite{S. C. Lee2}. A fundamental tone is generated by the microwave source and is sent to the superconducting sample through the directional coupler. Then the reflected harmonic response from the superconducting sample are shunted by the directional coupler and the selected harmonic response ($P_{3f}$) is analyzed by the spectrum analyzer.
   Inset (a) is a schematic diagram of the high RRR bulk Nb sample. The marked rectangle is the region where many single-fixed-point $P_{3f}$ measurements are performed. (b) shows a top view of the magnetic write head probe assembly on top of the high RRR bulk Nb sample; inset (c) is an optical reflection surface image of the bulk Nb sample.}
   \label{Setup_APL}
   \end{quote}
   \squeezeup
\end{figure}

In our first experiment, we measure the third harmonic scalar power $(P^{sample}_{3f})$ that is generated by the bulk Nb sample due to local harmonic excitation at a fixed location. The third-harmonic power is studied because it is sensitive to the extrinsic properties (i.e. associated with defects), almost always at a level far stronger than the intrinsic response \cite{Oates2007} and background levels. Fig. \ref{Setup_APL} shows the schematic microwave circuit for sensing small $P^{sample}_{3f}$ signals \cite{Tai2013}\cite{S. C. Lee1}\cite{S. C. Lee2} from a bulk Nb superconductor. A schematic diagram of the sample is also shown in the inset (a). Both the magnetic probe and superconductor are kept in a high vacuum cryogenic environment. The magnetic write head as shown in inset (b) of Fig. \ref{Setup_APL} is made by Seagate (part $\sharp$ GT5) and has a 100 nm wide magnetic gap. The microwave source generates an excitation wave (a fundamental tone) at frequency $f$ which is sent into the magnetic write head probe through a low pass filter (to eliminate harmonics) and directional coupler. The magnetic write head probe creates a localized and intense RF magnetic field on the surface of the bulk Nb sample, similar in spirit to previous studies of magnetic materials \cite{Schultz1996}\cite{Schultz1997}. The sample
responds by creating screening currents to maintain the Meissner
state in the material. These currents inevitably produce a
time-dependent variation in the local value of the superfluid
density, and this in turn will generate a response at harmonics of the driving
tone \cite{S. C. Lee2}\cite{Jeffries}. The generated harmonic signals are gathered by the magnetic
probe and returned to room temperature where they are high-pass
filtered to remove the fundamental tone $P_f$ and the second harmonic $P_{2f}$. The generated third harmonic signals $(P_{3f})$ are measured by the spectrum analyzer. The noise floor of the spectrum analyzer is -140 $\sim$ -145 dBm, depending on frequency and averaging time.

Note that the bulk Nb sample is a tensile testing sample that has an etched surface and came from the outer perimeter of the Nb ingot, where thermal strain due to ingot cooling was large. This sample has a high Residual-Resistance Ratio (RRR $\sim$ 200) (with $T_c=9.2$ $K$) and goes through the same material processing steps as a bulk Nb cavity \cite{Bieler2008}\cite{Bieler2010}\cite{Bieler2013}. Many single-fixed-point $P_{3f}$ measurements are performed on the essentially undeformed shoulder region inside the marked rectangle in inset (a) of Fig. \ref{Setup_APL}, where the surface is comparatively flat, permitting close probe/sample separation. An optical reflection image of the bulk Nb surface from the marked region is shown in inset (c) of Fig. \ref{Setup_APL}, showing surface topography resulting from etching used to remove 20 microns of the surface layer of side F in a tensile sample shoulder
region of welded sample FC described in \cite{Bieler2008}. A good thermal anchoring of the sample is required to ensure that the surface is superconducting, and a thermometer is placed on the bulk Nb top surface to monitor the temperature of the relevant surface of the bulk Nb. In order to control the height of the probe over the sample, we perform a resonant frequency measurement of the probe assembly as a function of probe/sample separation. The resonant frequency is used as the excitation frequency in the third harmonic response experiment. From our previous nonlinearity measurement and modeling on Nb thin films \cite{Tai2013}, the probe height is approximately several hundred $nm$ from the superconductor surface and creates a field parallel to the surface up to the level of the thermodynamic critical field of Nb at 0 K. The details of the probe height control are discussed in Ref. \cite{Tai_thesis}.
\begin{figure}
\centering
\includegraphics*[width=3.4in]{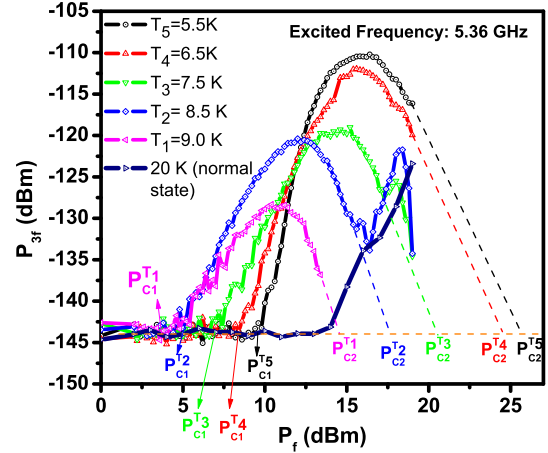}
\begin{quote}
\caption{The $P_{3f}$ dependence on $P_f$ at some specific temperatures for local microwave excitation of bulk Nb with the magnetic write head probe. Note the $T_c$ of the bulk Nb sample is 9.2 K. The $P_{3f}$ at $T=20$ $K$ results from the probe nonlinearity itself. $P_{c1}^{T_5}$ indicates the lower critical power, $P_{c1}$, at temperature $T_5$. $P_{c2}^{T_5}$ indicates the upper critical power, $P_{c2}$, at temperature $T_5$. $P_{c1}^{T_4}$, $P_{c1}^{T_3}$, $P_{c1}^{T_2}$, $P_{c1}^{T_1}$ and $P_{c2}^{T_4}$, $P_{c2}^{T_3}$, $P_{c2}^{T_2}$, $P_{c2}^{T_1}$ have analogous definition. Note that in the measurement, $T_5=5.5$ $K$, $T_4=6.5$ $K$, $T_3=7.5$ $K$, $T_4=8.5$ $K$, $T_1=9$ $K$.}
\label{NbP3fPfSet1}
\end{quote}
\squeezeup
\end {figure}

\begin{figure}
\centering
\includegraphics*[width=2.6in]{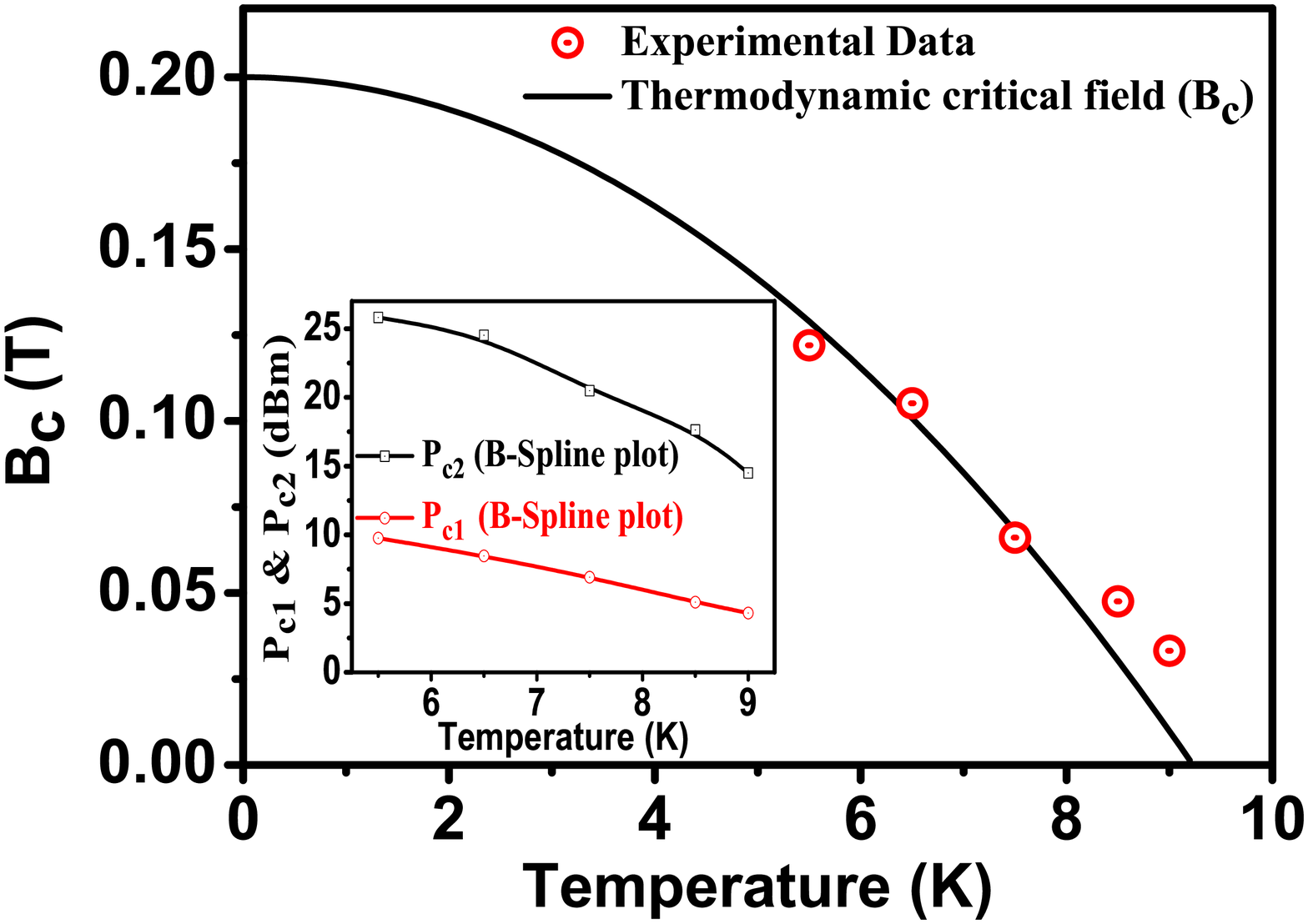}
\begin{quote}
\caption{A fit of the experimental temperature dependent thermodynamic critical field $B_{c}(T)$ (dot points) with the approximate equation (solid line): $B_{c}(T) \cong B_{c}(0K) \Big(1-(T/T_c)^2\Big)$. The dot points are the calculated $B_c$ from the experimental $P_{c2}$ in Fig. \ref{NbP3fPfSet1}. The inset is the temperature dependent $P_{c1}$ and $P_{c2}$ extracted
from the data on bulk Nb shown in Fig. \ref{NbP3fPfSet1}.}
\label{Pc2Bc2_fit}
\end{quote}
\squeezeup
\end {figure}

Figure \ref{NbP3fPfSet1} shows the representative measurements for the power dependence of $P_{3f}$ with respect to the fundamental input power $P_f$ at several fixed temperatures under 5.36 GHz excitation of one fixed location on bulk Nb. The curve measured at T=20 K indicates the probe third harmonic response $P_{3f}^{probe}$ on the surface of Nb in the normal state. The probe shows third harmonic response only at excitation powers of 15 dBm and above. This is because the magnetic write head is made of ferrite which has hysteretic characteristics and generates background nonlinearity \cite{Tai2011}. For measurement below $T_c$, all curves (at temperature $T_1,T_2,\cdots,T_5$) show a sharp $P_{3f}$ onset from the noise floor of the spectrum analyzer and then follow a continuous increase of nonlinearity until a turnover at high excitation power. After the turn over, the nonlinearity goes down until it approaches the curve of probe nonlinearity $P_{3f}^{probe}$. After that point, the measured nonlinearity oscillates around the curve of probe nonlinearity. The onset of the nonlinear response is temperature dependent and can be defined as a temperature dependent lower critical power $P_{c1}$, a minimum power to excite this nonlinearity above the noise floor. One can clearly see the relation of $P_{c1}$ for each temperature is $P_{c1}^{T_5}$ $>$ $P_{c1}^{T_4}$ $>$ $P_{c1}^{T_3}$ $>$ $P_{c1}^{T_2}$ $>$ $P_{c1}^{T_1}$. That is to say, the onset of nonlinearity requires higher excitation power at lower temperatures. In addition, after the turnover, all $P_{3f}(P_f)$ curves tend to decrease with increasing power, suggesting that we are suppressing superconductivity locally on the sample. The linear extrapolation of $P_{3f}$ to the noise floor can be defined as an upper critical power $P_{c2}$, which suggests that superconductivity will eventually be annihilated in high RF magnetic field. Clearly, the relation of $P_{c2}$ for each temperature is $P_{c2}^{T_5}$ $>$ $P_{c2}^{T_4}$ $>$ $P_{c2}^{T_3}$ $>$ $P_{c2}^{T_2}$ $>$ $P_{c2}^{T_1}$. In addition, the slope after each onset is increasing with decreasing temperature. The increase of slope, $P_{c1}$ and $P_{c2}$ with decreasing temperature are consistent with the temperature dependent critical fields of bulk Nb. A summary for both the temperature dependent $P_{c1}$ and $P_{c2}$ are plotted
together in the inset of Fig. \ref{Pc2Bc2_fit}.

An analysis of the temperature dependent $P_{c2}$ at $T_1,T_2,\cdots,T_5$ can be used to estimate the surface field excited by the magnetic write head probe. Generally, the relation between $P_{c2}$ and $B_c$ can be written as $P_{c2} (T)= k \left[B_c(T) \right]^2$, where $k$ is a constant relating the incident power in the probe to the RF magnetic field experienced by the Nb surface. The solid curve in Fig.  \ref{Pc2Bc2_fit} shows the temperature dependent thermodynamic critical field of bulk Nb which is approximated by $B_{c}(T)\cong B_{c}(0K) \left[1-\Big(\frac{T}{T_c}\Big)^2\right]$ with $B_c(0K)=0.20$ $T$ \cite{Padamsee} for the zero temperature thermodynamic critical field of Nb. Note that the value of $B_c(0K)$ depends on the quality of Nb such as RRR, degree of crystallization, annealing, processing etc. We assume that our high RRR Nb sample can achieve the BCS theoretical value at zero temperature. The points in Fig. \ref{Pc2Bc2_fit} are the experimental fit of $B_{c2}(T)$ by adjusting the value of $k$ to fit the experimental $P_{c2}$ data. From the field scale generated by a High Frequency Structure Simulator (HFSS) model of the probe \cite{Tai2013} and an analytical model of field distribution on the superconductor utilizing the Karlqvist equation \cite{Tai2012}, the surface field is on the scale of zero temperature thermodynamic critical field of Nb at a sample/probe separation of several hundred nanometers. When $k=25.6$ $W/T^2$, the model and the experimental $P_{c2}$ data have minimum deviation. From the fit, the concentrated RF-field on the bulk Nb surface reaches $\sim$ 120 $mT$. The same nonlinear measurement performed at other location shows the same shape of curve but different values of $P_{c1}$ and $P_{c2}$ with $\sim$ 5 dBm variation. The variation of $P_{c1}$ and $P_{c2}$ may due to different localized material properties or probe height variation for each measurement.
\begin{figure}
\centering
\includegraphics*[width=2.1in]{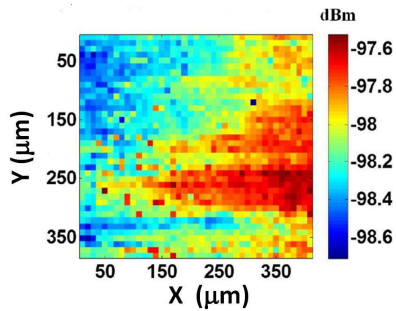}
\begin{quote}
\caption{A $P_{3f}$ image on a high quality MgB$_2$ thin film with thickness 200 $nm$.  This image is taken at T=15 K under 5.33 GHz and 16 dBm microwave excitation. Each pixel is 10 $\mu m$ by 10 $\mu m$ in size.}
\label{MgB2_image}
\end{quote}
\end{figure}
The second experiment is to perform a raster scan of the probe on a high quality MgB$_2$ thin film with thickness 200 nm made by hybrid physical-chemical vapor deposition on a sapphire substrate \cite{Xi}. The surface of this sample is much smoother than that of bulk Nb. Step motors are attached to the three axis translatable arm on the cryostat with a minimum step size 0.5 $\mu m$ \cite{Dinner}. This step size currently constrains the spatial resolution of our near field scanning probe microwave microscope. The detailed setup of the three axes stepping motor system is discussed in Ref. \cite{Tai_thesis} \cite{Dinner}.
The experiment is performed at T=15 K (below $T_c$ $\cong$ 35 K) under 5.33 GHz and 16 dBm microwave excitation. The excitation power is determined from the power dependence measurement for the maximum excitation power before the onset of the probe nonlinearity while in proximity to the MgB$_2$ sample in the normal state. The power dependent $P_{3f}(P_f)$ and temperature dependent $P_{3f}(T)$ at a fixed position are discussed in Ref \cite{Tai2012}. The scan direction is along the x direction first, and then the probe traces back to the beginning point, and an increase by one step in the y direction is made without lifting the probe. The step sizes for both x-axis and y-axis motors are 10 $\mu m$ in this raster scan.  Fig. \ref{MgB2_image} shows the $P_{3f}$ image on the MgB$_2$ sample. The $P_{3f}$ contrast shows stronger harmonic signal at the right part of the image but the difference of the position-dependent $P_{3f}$ is small, roughly 1 dB. This demonstrates that the MgB$_2$ is a homogeneous high quality film over a large scanning area and shows great potential for the inner surface coating on SRF cavities to increase the accelerating gradient in the future.

In summary, a clear reproducible nonlinear response signal from the surface of superconducting bulk Nb is obtained by the magnetic write head probe. The success of nonlinear excitation and localized suppression of superconductivity on high-RRR bulk Nb in the GHz frequency regime indicates the localized magnetic field from the magnetic write head probe is on the order of the thermodynamic critical field of Nb with $B_{surface}$ $\sim$ $10^2$ $mT$. This unique concentrated RF-field has great potential to identify the spatially-resolved electrodynamic properties of SRF cavity materials. The raster scan harmonic image of the MgB$_2$ thin film demonstrates the capability of large area scanning to identify the defects that cause break down of SRF cavities. \\

This work is supported by the US Department of Energy $/$ High Energy
Physics through grant $\#$ DESC0004950, and also by the ONR AppEl,
Task D10, (Award No.\ N000140911190), and CNAM. The work at MSU is supported by DOE-OHEP, contract number
DE-S0004222. The work at Temple University is supported by DOE under grant No. DE-SC0004410.

\end{document}